\documentclass[preprint,showpacs,preprintnumbers,amsmath,amssymb, superscriptaddress,longbibliography,nofootinbib]{revtex4-1}

\usepackage{textcomp}
\usepackage{makeidx}
\usepackage{amsmath}
\usepackage{subfig}
\usepackage{amssymb}
\usepackage{hyperref}
\usepackage{graphicx}
\usepackage{epstopdf}
\usepackage{color}
\usepackage{soul}

\begin{document}

\title{New models of d-dimensional black holes without inner horizon and with an integrable singularity}

\author{Milko Estrada\footnote{E-mail: milko.estrada@gmail.com}}\affiliation{Facultad de Ingeniería y Empresa, Universidad Católica Silva Henríquez, Chile}

\author{G. Alencar \footnote{E-mail: geova@fisica.ufc.br}}\affiliation{Departamento de F\'isica, Universidade Federal do Cear\'a, Caixa Postal 6030, Campus do Pici, 60455-760 Fortaleza, Cear\'a, Brazil}

\author{Tiago M. Crispim \footnote{E-mail: tiago.crispim@fisica.ufc.br}}\affiliation{Departamento de F\'isica, Universidade Federal do Cear\'a, Caixa Postal 6030, Campus do Pici, 60455-760 Fortaleza, Cear\'a, Brazil}

\author{C. R. Muniz\footnote{E-mail: celio.muniz@uece.br}}\affiliation{Universidade Estadual do Cear\'a (UECE), Faculdade de Educa\c c\~ao, Ci\^encias e Letras de Iguatu, 63500-000, Iguatu, CE, Brazil.}
\date{\today}

\begin{abstract}

Theoretically, it has been proposed that objects traveling radially along regular black holes (RBHs) would not be destroyed because of finite tidal forces and the absence of a singularity. However, the matter source allows the creation of an inner horizon linked to an unstable de Sitter core due to mass inflation instability. This inner horizon also gives rise to the appearance of a remnant, inhibiting complete evaporation. We introduce here a $d$-dimensional black hole model with Localized Sources of Matter (LSM), characterized by the absence of an inner horizon and featuring a central integrable singularity instead of an unstable de Sitter core. In our model, any object tracing a radial and timelike world-line would not be crushed by the singularity. This is attributed to finite tidal forces, the extendability of radial geodesics, and the weak nature of the singularity. Our LSM model enables the potential complete evaporation down to $r_h=0$ without forming a remnant. In higher dimensions, complete evaporation occurs through a phase transition, which could occur at Planck scales and be speculatively driven by the Generalized Uncertainty Principle (GUP). Unlike RBHs, our model satisfies the energy conditions. We demonstrate a linear correction to the conventional area law of entropy, distinct from the RBH's correction. Additionally, we investigate the stability of the solutions through the speed of sound.
\end{abstract}

\maketitle

\section{Introduction}
The detection of gravitational waves resulting from the collision of rotating black holes has propelled these mysterious objects to the forefront of gravitational studies \cite{LIGOScientific:2016aoc}. Black hole solutions typically exhibit a central singularity where metric and curvature invariants diverge. It is well known that infinite tidal forces located near the singularity of a black hole can cause unlimited stretching of an object, a phenomenon known as spaghettification. In this regard, there are several studies in the literature where the disintegration of stars due to their interaction with tidal forces is proposed (see the introduction of \cite{Hong:2020bdb} and references therein). To address this problem, one approach involves defining the energy-momentum tensor in a specific way, resulting in the formation of regular black holes (RBH). RBHs are characterized by having finite values for curvature invariants and finite tidal forces throughout their spacetime \cite{Lima:2020wcb}

Localized sources of matter (LSM) have been a widely employed instrument in the literature to characterize the distribution of matter in the energy-momentum tensor of RBHs. Examples of $4D$ RBH models with LSM include the Dymnikova model \cite{Dymnikova:1992ux} and a variant of the Hayward model \cite{DeLorenzo:2014pta}. Beyond four dimensions, $d$--dimensional RBH models with LSM have been proposed, where physical properties may vary with the number of extra dimensions \cite{Aros:2019quj,Estrada:2019qsu, Estrada:2023cyx}. Additionally, a $(2+1)$ RBH model with LSM has been discussed \cite{Estrada:2020tbz, Hendi:2022opt}. However, in some cases, such as the $(2+1)$ model, LSM behavior may conflict with predictions from nonlinear electrodynamics sources \cite{Maluf:2022jjc}. 

The primary motivation behind RBH solutions is to replace the central singularity with a de Sitter core located in the vicinity of an inner horizon. Thus, RBH solutions typically incorporate an inner horizon, where predictability breaks down, and several physical issues arise as well \cite{Ovalle:2023vvu}. In this context, concerns persist regarding the stability of de Sitter cores due to their location. Instabilities localized at the inner horizon, leading to mass inflation and a breakdown of fundamental physics, have been explored \cite{Poisson:1990eh,Brown:2011tv}. Proposed conditions for correcting instability at the inner horizon have been discussed \cite{Carballo-Rubio:2022kad,Bonanno:2020fgp}. However, the theoretical viability of cores in RBHs, given their instability, remains an ongoing issue in physics \cite{Carballo-Rubio:2018pmi,DiFilippo:2022qkl}. Recently, reference \cite{Carballo2023} claims that the instability of cores in RBHs is mandatory due to the relevance of the mass inflation instability for RBHs of astrophysical interest.

On the other hand, reference \cite{Lukash:2013ts} suggests that the existence of a central singularity leads to finite tidal forces along the world line of the matter flow, thereby preventing the destruction of objects approaching it. An integrable singularity is one where the energy-momentum tensor and the Ricci scalar diverge, while their volume integrals have a finite value. Specifically, this reference demonstrates that a $4D$ energy density that behaves like $\rho \sim r^{-2}$ near the origin ensures that the tidal forces remain finite along any radial geodesic, contrasting with typical singular black holes where tidal forces are infinite near the singularity \cite{Lima:2020wcb}. Thus, an integrable singularity could be seen as an alternative to RBHs to prevent infinite tidal forces from destroying an object as it heads towards the center of a black hole.

Motivated by the fact that an integrable singularity akin to the one described in the preceding paragraph is associated with finite tidal forces in its vicinity, and considering the array of physical issues arising from the existence of a Sitter core located near the inner horizon, reference \cite{Casadio:2023iqt} proposed a specific model of four-dimensional black holes with a central integrable singularity and without the presence of an inner horizon. In this model, the energy density and mass function behave near the origin as $\rho \sim r^{-2}$ and $m \sim r$, respectively. This approach allows radial geodesics to extend up to the central singularity. Additionally, this study suggests a speculative quantum perspective wherein black holes could arise from quantum matter described by wavefunctions. The energy density, similar to the wavefunction of the static matter source, follows $\rho \propto |\psi|^2$, where $\psi$ must be integrable to ensure finite probability densities. See also \cite{Casadio:2023ymt}. While speculative, this quantum viewpoint requires further investigation to establish its consistency with gravitational theory.

To address the problems related to the appearance of an inner horizon in the presence of matter in the energy-momentum tensor, and the existence of infinite tidal forces that destroy objects approaching the singularity radially,  in this work we will provide a $d$-dimensional black hole model with LSM in the energy-momentum tensor that lacks an inner horizon and has an integrable singularity. To do this, we will determine the constraints that must be satisfied by generical LSMs and the black hole metric, both in four-dimensional and extra-dimensional models, to ensure the existence of an integrable singularity and to prevent the formation of an internal horizon associated with an unstable Sitter core. We will test the conditions for the tidal forces to be finite in the radial vicinity of such a singularity and for radial geodesics to be extendable to it. Furthermore, we will test if our type of central integrable singularity is weak in nature \cite{Tipler:1977zza,Nolan:2000rn}, and thus, if any object following a radial and timelike world-line would not be crushed by this gravitationally weak singularity. We will discuss the speculative analogy with quantum effects presented in references \cite{Casadio:2023iqt,Casadio:2023ymt}. Additionally, we will propose a toy model as a test of proof for the energy density and metric tensor that satisfy the conditions described in this paragraph.

On the other hand, a well-known phenomenon in the evaporation processes of RBHs involves the contraction of the event horizon ($r_h$) until it coincides with the inner horizon ($r_h = r_{in}$), forming what is termed an extremal black hole. Once this state is reached, evaporation halts, and the black hole's temperature decreases to zero, ceasing the emission of heat $Q$ into the environment. Consequently, the final state post-evaporation is a black remnant at zero temperature. This halt in the thermal emission prevents RBHs from completely evaporating (i.e., where the event horizon contracts to $r_h=0$). Regarding this, it was conjectured in reference \cite{Estrada:2023cyx} that for higher-dimensional RBHs, the black remnant radius could be on the order of the Planck length. This phenomenon could be linked to the emergence of quantum effects at such scales. It is argued that the Generalized Uncertainty Principle (GUP) at Planck scales should inhibit total evaporation, analogous to how the uncertainty principle prevents the collapse of the hydrogen atom at atomic scales \cite{Adler:2001vs}. Additionally, in some scenarios where the heat capacity remains negative, such as in the vacuum solution of Schwarzschild, achieving complete evaporation until the event horizon vanishes necessitates infinite temperature growth. It's worth mentioning that there are studies in the literature suggesting that black holes should evaporate completely \cite{Piran:1993tq,Li:2021mjp,Arefeva:2022guf}. In our proposed model without an inner horizon, exploring the dynamics of evaporation processes becomes intriguing, especially considering the impossibility of forming a remnant with $r_h=r_{in}$. Understanding phase transitions is crucial for comparing the final stages of evaporation processes with other black hole solutions. It is pertinent to investigate whether complete evaporation until the event horizon vanishes is feasible without an inner horizon and explore potential outcomes in the presence of extra dimensions. Hence, our study also will delve into these issues concerning our proposed model.

Furthermore, it is well known that energy conditions can be violated by matter associated with RBH \cite{Zaslavskii:2010qz}. Therefore, in this work, we will also investigate whether our type of LSM can satisfy the energy conditions. Additionally, we will conduct an analysis of solution stability based on calculations of fluid sound velocity.

In RBHs, the presence of matter fields in the energy-momentum tensor alters the structure of the first law of thermodynamics. These modifications have led to corrections in the area law of entropy, with references offering logarithmic and exponential/Ei function corrections \cite{Estrada:2020tbz, Maluf:2022jjc, Estrada:2023cyx, Morales-Duran:2016jqt, Singh:2022xgi}. In our study, we also investigate the implications of our type of matter on the first law of thermodynamics and explore the induced corrections to the area law of entropy.

\section{Our model of $d$-dimensional black hole with an integrable singularity and without an internal horizon}\label{Restricciones}

In this section, we will define our generic model of a black hole with an integrable singularity and without an internal horizon. To achieve this, we will establish the necessary constraints that both LSMs and the metric tensor must satisfy to ensure the presence of an integrable singularity and to prevent the formation of an inner horizon associated with an unstable de Sitter core. We will test the conditions for the tidal forces to be finite in the radial vicinity of such a singularity and for radial geodesics to be extendable to it. Furthermore, we will test if our type of central integrable singularity is weak in nature \cite{Tipler:1977zza,Nolan:2000rn}, and thus, if any object following a radial and timelike world-line would not be crushed by this gravitationally weak singularity. Additionally, we will revisit the speculative analogy between the energy density and the quantum wavefunction presented in reference \cite{Casadio:2023iqt}. It is important to note that this analogy is purely speculative, and establishing a fully consistent relationship between gravitational and quantum theories requires further in-depth study. The Einstein field equations are given by:
\begin{equation}
    G^\mu_\nu=8\pi G_d T^\mu_\nu,
\end{equation}
where $G^\mu_\nu, T^\mu_\nu$, and $G_d$ represent the Einstein tensor, the energy-momentum tensor and the higher dimensional Newton constant. 

We study the following $d$--dimensional static spherically symmetric space--time:
\begin{equation} \label{elementodelinea1}
ds^2 = -f(r) dt^2 + f(r)^{-1} dr^2 + r^2 d\Omega_{d-2}
\end{equation}
where $d\Omega_{d-2}$ corresponds to the transversal section of a $(d-2)$ sphere. Furthermore, we study the following form for the energy-momentum tensor:
\begin{equation}\label{TensorEM}
    T^{\mu}_{\nu} =  \textrm{diag}(-\rho(r), p_r(r),p_\theta(r) ,p_\phi(r), ...).
\end{equation}
where $\rho,p_r$ are the energy density and the radial pressure and where $p_\theta ,p_\phi,...$ are the $(d-2)$ angular coordinates.
Due to the spherical symmetry in all $(d-2)$ angular coordinates, it is satisfied that $p_\theta=p_\phi=...=p_t$. Furthermore, from the line element form \eqref{elementodelinea1} is satisfied that $\rho=-p_r$. Thus, the $(t,t)$ and $(r,r)$ components of the equations of motion are given by:
\begin{equation} \label{eqmotion}
    \left ( r^{d-3} (1-f)  \right )' = \frac{16 \pi G_d}{d-2} r^{d-2}\rho,
\end{equation}
where $'$ indicates derivation with respect to the radial coordinate. The solution is:
\begin{equation} \label{Solucion1}
    f(r)= 1- \frac{2 G_d m(r)}{r^{d-3}},
\end{equation}
where 
\begin{equation} \label{funcionDeMasaCompleta}
m(r) = \frac{8\pi}{(d-2) \Omega_{d-2}} \bar{m}(r)
\end{equation}
and where
\begin{equation} \label{FuncionDeMasa}
    \bar{m}(r)= \int \Omega_{d-2} r^{d-2} \rho dr
\end{equation}
On the other hand, tangential pressures can be computed using the conservation equation: 
\begin{equation} \label{Conservacion}
    - \frac{r}{d-2} \rho' - \rho= p_t
\end{equation}

In this section, we will determine the necessary conditions that all types of energy density with LSM and the mass function must satisfy, both for the $4D$ case and the extra-dimensional case, for 
to obtain a black hole solution with an integrable singularity and to prevent the formation of an internal horizon associated with an unstable Sitter core.
We consider the general behavior $\rho \sim \bar{C} r^{-N}$ near the origin, with $\bar{C}>0$ in order $\rho>0$. Next, we study some possibilities, detailed below:

\begin{enumerate}
    \item Disregarding the RBH solution with inner horizon \label{InnerHorizon}
    
   Testing the condition $\rho \sim \bar{C} r^{-N}$, near the origin, in equation  \eqref{FuncionDeMasa}
\begin{equation} \label{CondicionMasa}
\bar{m} \sim C r^{d-1-N}
\end{equation}
It is known that $N=0$ leads to regular Kretshamn and Ricci scalars. However, the behavior near the origin of the last factor of equation \eqref{Solucion1}
\begin{equation}
    \frac{2 G_d m(r)}{r^{d-3}} \sim \frac{C r^{d-1-N}}{r^{d-3}}= C r^{2-N}
\end{equation}
thus, near the origin, the behavior is:
\begin{equation} \label{core}
 f \approx 1- C r^{2-N}
\end{equation}

As was mentioned, several studies in the literature (see, for example, \cite{Dymnikova:1992ux}) have demonstrated that a finite value of energy density near the origin, denoted as $\rho \sim r^0$ with $N=0$, causes the latter to act as a positive cosmological constant at this location. This results in a de-Sitter-like behavior near the origin, where the signature of the metric exterior to the event horizon has been restored, implying the presence of an inner horizon close to the de--Sitter core, given by the positive value of $r_{in}$ such that $f \approx 1- Cr_{in}^2=0$ in equation \eqref{core} with $N=0$. Thus, to prevent the presence of an inner horizon for small values of the radial coordinate, {\it i.e.}, to ensure that equation \eqref{core} does not yield a value of the radial coordinate such that $f=0$, the only allowed value is

\begin{equation} \label{Condicion2}
N=2 \Rightarrow  \rho \sim \bar{C} r^{-2} \mbox{ \,\,\, near the origin}
\end{equation}

From conditions \eqref{CondicionMasa} and \eqref{Condicion2}, the behavior of the mass function near the origin varies for different numbers of dimensions $d$:

\begin{equation} \label{CondicionMasaFinal}
m \sim C r^{d-3}
\end{equation}

Furthermore, from equation \eqref{core}, which takes the form 
\begin{equation} \label{funcionCore}
  f \approx 1-C  
\end{equation}
at the origin, it must be satisfied that $C \ge 1$ to avoid recovering the exterior signature once it has crossed the event horizon. This is because $f<0$ immediately after crossing the mentioned event horizon.

Hence, an energy density that behaves near the origin as $\rho \sim C r^{-2}$ avoids the presence of an internal horizon associated with an unstable Sitter core. The mass function must behave as $m \sim r^{d-3}$ near the origin, which reduces to Casadio's result $m \sim r$ for $d=4$.

\item Integrability of the energy-momentum tensor: \label{EnergiaMomentum}

As mentioned in the introduction, one of the conditions associated with the integrable singularity existence is that the volume integral of the energy-momentum tensor near the origin has a finite value. In this connection, Casadio et al. consider that the probability is related to the mass function \eqref{FuncionDeMasa}, where it is assumed that the energy density is analogous to the wave function. Let us consider the case with extra dimension and find a more general constraint. First of all, to be integrable close to the origin, we must have:
    
  \begin{equation} \label{Condicion1}
    r^{d-2} \rho \sim \bar{C} r^{d-4}= \mbox{finite} \Rightarrow d \ge 4
\end{equation}

 For every value of $d \ge 4$, the energy density must behave as $\rho \sim r^{-2}$ near the origin. Thus, we point out that, the number of dimensions of the space-time must be such that $d \ge 4$. Therefore, the above strategy is not valid for the $(2+1)$ case. 

On the other hand, from equation \eqref{Conservacion}, it is direct to check that the tangential component of the energy-momentum tensor also behaves as $r^{-2}$ near the origin. Thus, the volume integral of the energy-momentum tensor components is also finite near the origin for $d \ge 4$.

\item Well posed physical situation: \label{WellPosed}

The energy density must be a positive and decreasing radial function such that the factor $r^{d-2} \rho$ is free of singularities everywhere.

On the other hand, the energy density must approach zero at infinity to ensure well-defined asymptotic behavior. This implies that the mass function reaches its finite maximum value at infinity
\begin{align}
\displaystyle &\lim_{r \to \infty} \rho(r) =0 \label{AsintotaRho} \\
\displaystyle &\lim_{r \to \infty} m(r) = M >0 \label{AsintotaMasa}
\end{align}

 \item Integrability of Ricci scalar \label{Ricci}

 The higher dimensional Ricci invariant \cite{Gurses:1995fw} near the origin behaves as:
\begin{equation} \label{RicciOrigen}
    R \sim \frac{(d-2)(d-3)C}{r^2}
\end{equation}
Thus, under their assumptions, the volume integral $\sim R r^{d-2}$ is finite near the origin for $d \ge 4$.

\item Finite radial Geodesics \label{Geodesicas}

To determine whether radial geodesics can extend up to the central singularity, we test whether a radial timelike geodesic remains finite near the origin. The angular part is:

\begin{equation}
 \displaystyle   d\Omega_{D-2}= d\theta^2_1 + \sum_{j=2}^{D-2} d\theta^2_j \left ( \prod_{k=1}^{j-1} \sin^2\theta_k  \right) 
\end{equation} 
with $j=1...D-3$. We consider a radial movement described by $\theta_1 = \theta_2 = \ldots = \theta_{D-3} = \pi/2$ and $\theta_{D-3}$ constant. Thus, any timelike radial geodesic is given by \cite{Vandeev:2022xqk}:
\begin{equation} \label{geodesica1Radial}
\dot{r}^2 + V_{\text{eff}} = \frac{E^2}{2}
\end{equation}
where
\begin{equation} \label{geodesica2Radial}
    V_{\text{eff}} = \frac{1}{2} f(r)
\end{equation}
where, since the function \eqref{funcionCore} is finite at the origin, we can deduce that any timelike radial geodesic motion can be extended up to the central singularity.

{\bf Note}: We can observe that, if we were to begin our analysis with the fact that the dependence of the Ricci tensor $R\sim r^{-2}$ near the origin, equation \eqref{RicciOrigen} , via the trace of the Einstein equations $(1-d/2)R=T=(d-2)p_t$, consequently, the tangential pressure possesses the same dependence, and through equation \eqref{Conservacion} consequently the energy density shares this dependence as well. This latter, via equation \eqref{FuncionDeMasa}, implies that the mass function behaves as equation \eqref{CondicionMasaFinal} near the origin. Finally, this implies that the function $f$, equation \eqref{funcionCore}, is finite near the origin. Thus, the effective potential, equation \eqref{geodesica2Radial}, is finite. Therefore, for our model, the fact that the Ricci tensor has this behavior near the origin implies that a timelike radial geodesic can extend to the singularity.

\item Finite tidal forces: \label{FuerzasTidales}

We are interested in the tidal forces of a non-rotating radial geodesic. The geodesic deviation, which governs how quickly two particles separated by the spatial vector $\xi^a$ accelerate relative to each other, reads \cite{Vandeev:2022xqk}:

\begin{align}
    \ddot{\xi}^r &= \left( -\frac{f''}{2} \delta_1  \right) \xi^r \\
     \ddot{\xi}^a &= \left( -\frac{f'}{2r} \delta_1  \right) \xi^a \\
     \ddot{\xi}^{\theta_{D-2}} &= \left( -\frac{f'}{2r} \delta_1  \right) \xi^{\theta_{D-2}}
\end{align}
where the dot indicates the derivative with respect to the affine parameter and where $a = (\theta_1, \ldots, \theta_{D-3})$ correspond to polar angular components and the temporary component is trivial  $\ddot{\xi}^t =0$.

Thus, we can deduce that generically the function $f$ and its second derivative must be finite near the origin, while its first derivative should not vanish with a power higher than $r^1$. These conditions are similar to those in the four-dimensional case \cite{Lukash:2013ts}. For our case, where $\rho \sim \bar{C} r^{-2}$ and where $f=$ constant near the origin, it is straightforward to check that the first and second derivatives of $f$ vanish, and thus the tidal forces are finite for a timelike radial geodesic. On the other hand, as indicated in the reference \cite{Lukash:2013ts}, tidal forces in transverse directions would not destroy an object moving radially towards the singularity .

\item Strength of the central singularity \label{Tipler}

Firstly, in 1977, Tipler \cite{Tipler:1977zza} stablished a theorem to determine whether any object following its world-line will inevitably be crushed by a gravitationally strong singularity. Afterward, conditions to determine whether a singularity is weak or strong in a spherically symmetric spacetime were established in reference \cite{Nolan:2000rn}. In particular, it is established that:  if a radial causal geodesic $\gamma$, without angular momentum, terminates in a deformationally weak central singularity, then along $\gamma$, the value of:

\begin{equation}
 \displaystyle \lim_{\lambda \to 0} x(\lambda)= \lim_{\lambda \to 0}  r(\lambda) \int_{\lambda_1}^{\lambda} \frac{d\bar{\lambda}}{r(\bar{\lambda})^2}
\end{equation}
is finite and nonzero. Furthermore, it is states that there exist $c_0>0$ such that:
\begin{equation}
    r(\lambda) \sim c_0 \lambda \mbox{\,\,\, for\,\, } \lambda \to 0
\end{equation}
See an application of this theorem for a timelike radial geodesic without angular momentum in reference \cite{Maeda:2011px}. 

From equations \eqref{geodesica1Radial}, \eqref{geodesica2Radial}, and \eqref{funcionCore}, it is straightforward to verify that under our imposed constraints, the geodesic can be expressed as $r(\lambda) \sim c_0 \lambda$. Consequently, we can infer that our type of central integrable singularity is weak in nature. Therefore, any object following a radial and timelike world-line would not be crushed by this gravitationally weak singularity.

\end{enumerate}

In this section, we have enunciated the essential constraints that both LSMs and the metric tensor of our model must satisfy to allow for an integrable singularity and to prevent the formation of an internal horizon associated with an unstable Sitter core. We have examined the conditions ensuring finite tidal forces in the radial vicinity of such a singularity and the extendability of radial geodesics to it. Furthermore, we have deduced that our type of central integrable singularity is weak in nature. Therefore, any object following a radial and timelike world-line would not be crushed by this gravitationally weak singularity.

It is worth mentioning that the above prescription can be used as a recipe for constructing several new $4D$ and extra-dimensional solutions with LSM in the energy-momentum tensor, featuring a central integrable singularity, without the presence of an inner horizon, all while maintaining the speculative analogy between the energy density and the quantum wave function proposed in reference \cite{Casadio:2023iqt}. 

However, the absence of a full solution does not allow for studying stability and thermodynamic properties.  In the next section, we will propose a model with all the desired properties to achieve this.

\section{Our proposed model}

In this section, we will construct a full-density and analytical solution that recovers Casadio's behavior near the origin for $d=4$. As shown in the last section, for any number of dimensions, we must have that close to the origin $\rho\sim r^{-2}$. 

Let us begin with the second of the above conditions. If we consider that $m(r)=P(r)/Q(r)$, where $P$ and $Q$ are polynomials, they must be of the same order to satisfy $\displaystyle \lim_{r \to \infty} P(r)/Q(r) = M$. Next, we must have $P/Q \sim r^{d-3}$ from the behavior close to the origin. Finally, to avoid the inner horizon, we must have that 
\begin{equation}
 P(r)-r^{d-3}Q(r)=0   
\end{equation}
has only one solution. The last condition can be simplified if we choose $P\propto r^{d-3}$ and $Q=A+Br^{l}$. Using the last conditions, we find that $Q=A+r^{d-3}$ and
\begin{equation} \label{funcionMasa}
m(r)=\frac{Mr^{d-3}}{A+r^{d-3}}
\end{equation}
So, with this mass function, the solution is given by the equation \eqref{Solucion1}. Let us see if this gives the desired properties for $\rho$. This can be easily obtained and give
\begin{equation} \label{densidadEnergia}
\rho = \frac{d-2}{8\pi} \frac{A(d-3)M}{r^2(A+r^{d-3})^2} 
\end{equation}
where $A>0$ is a constant of units $\ell^{d-3}$, where $\ell_p$ corresponds to Planck units, and where $M$ corresponds to the mass parameter, which has units of $\ell_p^{-1}$. Thus, the energy density has units of $\ell_p^{-d}$ \cite{Aros:2019quj}.

It is worth to mention that replacing the energy density \eqref{densidadEnergia} in the equation \eqref{FuncionDeMasa}, using $\bar{C}=\frac{(d-2)\Omega_{d-2}}{8 \pi}M$ as constant of integration, and after this, replacing in equation \eqref{funcionDeMasaCompleta} we also obtain the mass function \eqref{funcionMasa}.

We can observe that near the origin, the mass function \eqref{funcionMasa} behaves according to equation \eqref{CondicionMasaFinal}, and the energy density behaves as $\rho \sim r^{-2}$. We can verify that our model of energy density and mass function satisfies the conditions \ref{InnerHorizon}, \ref{EnergiaMomentum}, \ref{WellPosed}, and \ref{Ricci} of Section II.

\begin{figure}[!h]
    \centering
    \begin{minipage}{0.5\linewidth}
        \centering
        \includegraphics[width=1.0\textwidth]{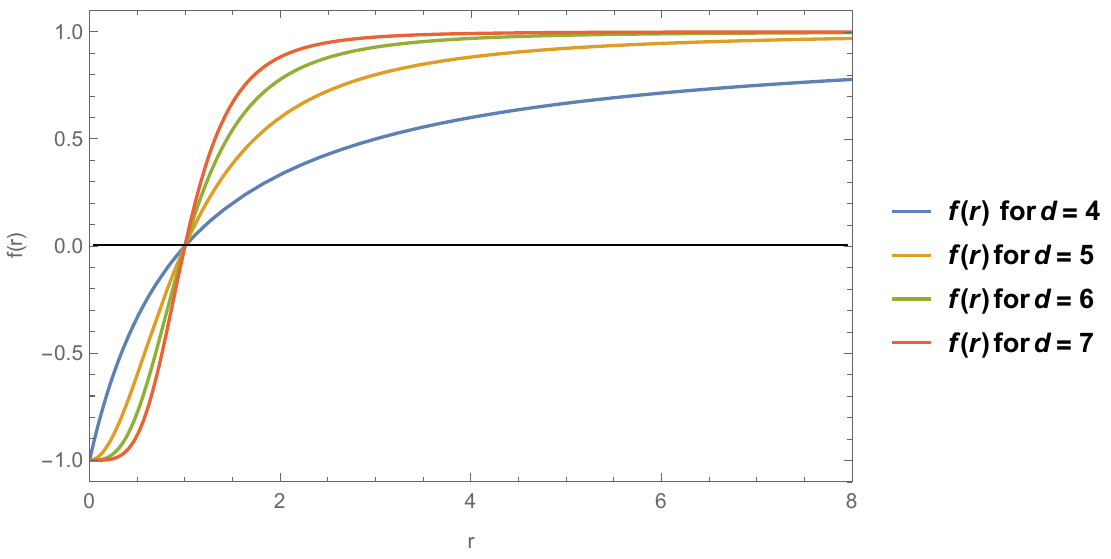}
        \label{Mplot}
    \end{minipage}\hfill
     \caption{$f(r)$ for $d=4,5,6,7$ with $A=M=1$ }
    \label{PlotHorizontes}
\end{figure}

It is worth mentioning that,  taking $G_d=1 \ell_p^{d-2}$ \cite{Aros:2019quj}, the solution \eqref{Solucion1} with the mass function \eqref{funcionMasa} has only one positive root $f=0$ given by:
\begin{equation}
    r_h = \left ( 2M-A  \right )^{1/(d-3)}
\end{equation}
It is easy to verify that, in order not to recover the exterior signature beyond the event horizon, once the latter has been crossed, it must be satisfied that $2M \ge A$.  It is worth noting that for odd $d$, the solution to $f=0$ given by $-( 2M-A )^{1/(d-3)}$ is discarded as its value is negative. In the figure \ref{PlotHorizontes} we see a generic behavior of our solution.

It is worth mentioning that our solution exhibits a finite value near the origin, taking the form of equation \eqref{funcionCore}, where, in our case, $C=2M/A$. In this context, it is straightforward to verify that the first and second derivatives of $f$ vanish near the origin. Consequently, our solution also satisfies the conditions \ref{Geodesicas}, \ref{FuerzasTidales}, and \ref{Tipler} outlined in Section II.

Thus, our solution lacks the presence of an internal horizon and, therefore,  the presence of an unstable core, as described in the introduction. Because our model satisfies the previously mentioned constraints, we can assert that our model leads to an integrable central singularity. Beyond this, we can now study stability and the thermodynamic aspects. 

\section{Thermodynamics analysis}

\subsection{\bf The total Energy}

To compute the total energy of the solution, we first employ the Komar formula \cite{Komar:1958wp}, which represents the conserved charge associated with the invariance of the action principle under diffeomorphisms. For our spacetime, the Komar formula is expressed as follows:

\begin{align}
   E \propto & \displaystyle K(\xi)= \nonumber \\
   =&\lim_{r \to \infty} \frac{1}{16 \pi G_d} \frac{d}{dr} f(r) r^{d-2} \int d\Omega_{d-2} \nonumber \\
   =&   (d-3)\frac{\bar{M}}{2},
\end{align}
where $\xi$ is a timelike Killing vector and where $\bar{M}=\frac{\Omega_{d-2}}{4\pi}M$. After regularization, which involves incorporating boundary terms into the action \cite{Aros:1999kt}, it is obtained that:
\begin{equation}
    E= 2 K(\xi) = (d-3) \bar{M}
\end{equation}
Thus, the value of the energy depends on the value of the dimensions, and for $d=4$ is recovered the value $E=M$.

\subsection{ \bf Correction to entropy and the first law of thermodynamics}

It is well known that the structure of the first law of thermodynamics is modified in the presence of matter fields in the energy-momentum tensor. For example, in Regular Black Hole Solutions (RBHs) where matter fields are present, this problem has been addressed using different approaches. One approach involves modifying the definition of internal energy \cite{Estrada:2020tbz,Maluf:2022jjc,Estrada:2023cyx}. Another approach consists of evaluating the first law as $dM = TdS$, which has led to corrections to the area's law in the entropy for RBHs. For instance, in reference \cite{Morales-Duran:2016jqt}, there is a logarithmic correction, and in reference \cite{Singh:2022xgi}, there is a correction provided by the exponential and the Ei function. 

Motivated by the fact that our proposed type of matter leads to black holes with different properties than RBHs, in this work, we test the corrections to the area law, following conditions $f(r_h,M,A)=0$ and $\delta f(r_h,M,A)=0$, which can be viewed as constraints on the parameter space, thus:
\begin{equation}
    0 = \frac{\partial f}{\partial r_+} dr_+ + \frac{\partial f}{\partial M} dM + \frac{\partial f}{\partial A} dA
\end{equation}
which give rise to:
\begin{equation}
dM= T \cdot d\left (  \frac{2}{d-2} \pi r^{d-2} + 2 \pi A r_h  \right ) - dA
\end{equation}
which is analogous to the first law of thermodynamics given by $dU=dM=TdS-dW$ and where the temperature is given by equation \eqref{TemperaturaMaster}. Thus the parameter $M$ is related to the system energy (as we saw previously via conserved charges) and the parameter $A$ could be related to the work done by the system. Thus, the presence of the parameter $A$ contributes to a decrease in the amount of energy in the system, which is usually given by $dM=TdS$, because the system releases an amount of energy to the environment in the form of work $dW=dA$. Thus, the entropy is given by
\begin{equation} \label{EntropiaMaster}
    S= \frac{2}{d-2} \pi r^{d-2} + 2 \pi A r_h
\end{equation}
Here we notice that the first term is proportional to the usual area law, $\sim r^{d-2}$, whereas the second term represents a linear correction. Thus, our proposed type of matter leads to corrections that differ from those previously mentioned for RBHs. 

In a four-dimensional scenario, the mentioned correction is also proportional to the square of the area of a 2-sphere, $\sim \sqrt{area}$. This could be related to the first-order modification (on an arbitrary variable $\alpha$) of the Generalized Uncertainty Principle (GUP) as referenced in \cite{Majumder:2011xg}, which also leads to a correction in entropy proportional to $\sim \sqrt{area}$. Furthermore, reference \cite{Amelino-Camelia:2005zpp} discusses coefficients that modify the GUP, where one of them also leads to a correction proportional to $\sim \sqrt{area}$. Therefore, speculatively, our linear correction could be associated with quantum effects through GUP corrections in four-dimensional scenarios. However, this latter one requires further in-depth future work.

\subsection{ \bf Temperature}
For our spacetime, the temperature is given by:
\begin{equation} \label{TemperaturaMaster}
    T= \frac{1}{4\pi} \frac{\partial f}{\partial r} \Big |_{r=r_h}
\end{equation}

Evaluating in our model whose mass function is \eqref{funcionMasa}
\begin{equation} \label{Temperatura}
    T= \frac{d-3}{4\pi} \frac{r_h^{d-4}}{A+r_h^{d-3}}
\end{equation}

\paragraph{\bf Case $d=4$:} \label{AnalisisTemperatura4D}

\begin{figure}[!h]
    \centering
    \begin{minipage}{0.5\linewidth}
        \centering
        \includegraphics[width=1.0\textwidth]{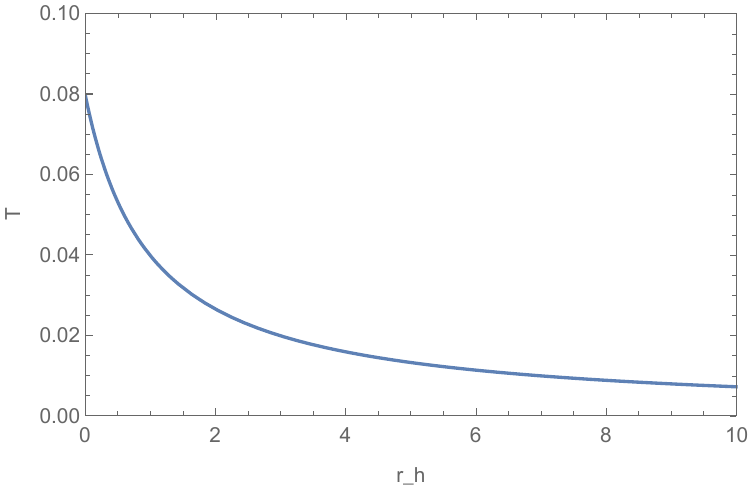}
        \label{Mplot}
    \end{minipage}\hfill
     \caption{$T$ for $d=4$ with $A=1$ }
    \label{PlotTemperaturad4}
\end{figure}

Firstly, we can point out that for $d=4$, the temperature takes the form $T= 1/ \left ( 4 \pi (A+r_h) \right)$. Thus, $T$ is always a decreasing function, and its negative derivative is given by $dT/dr_h=- 1/ \left ( 4 \pi (A+r_h)^2 \right)$. We can observe the graphical behavior of the temperature for $d=4$ in figure \ref{PlotTemperaturad4}.

Furthermore, unlike the Schwarzschild vacuum solution, where $T \sim r_h^{-1}$, and thus, for $r_h \to 0 \Rightarrow T \to \infty$, in this case, for a vanishing event horizon, the temperature reaches a finite value given by $T= \frac{1}{4\pi A}$. We will discuss the consequences of this property on the evaporation process below.

\begin{figure}[!h]
    \centering
    \begin{minipage}{0.5\linewidth}
        \centering
        \includegraphics[width=1.0\textwidth]{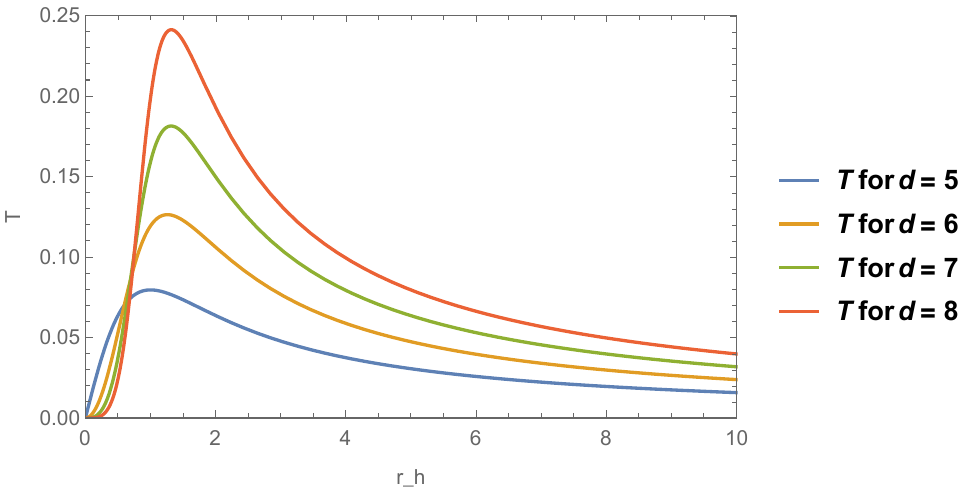}
        \label{Mplot}
    \end{minipage}\hfill
     \caption{$T$ for $d=5,6,7,8$ with $A=1$ }
    \label{PlotTemperaturaD}
\end{figure}

\paragraph{\bf Extra--dimensional case :} It is direct to check that, equation \eqref{Temperatura} has a local maximum at the following value of the event horizon:
\begin{equation} \label{valorCritico}
    r_h^{cri}= \left ( (d-4) A  \right)^{1/(d-3)}
\end{equation}
where the derivative $dT/dr|_{r=r_h^{cri}}$ vanishes. We can see a generic behavior of the temperature for $d>4$ in figure \ref{PlotTemperaturaD}. We can observe that for values lower than $r_h^{cri}$, the temperature is an increasing function, whereas for values greater than $r_h^{cri}$, the temperature behaves as a decreasing function. We can verify from equation \eqref{valorCritico} that a local maximum exists only for higher dimensions $d>4$. Below, we will discuss the consequences of these characteristics on the evaporation process.

\subsection{\bf Heat Capacity and Thermodynamics evolution}

In our investigation, heat capacity is defined as :
\begin{equation} \label{CalorEspecifico}
    C= \frac{dQ}{dT}=T \frac{dS}{dT} = T \left ( \frac{\partial S}{\partial r_h} \right) \left ( \frac{\partial T}{\partial r_h} \right)^{-1}
\end{equation}
The heat capacity will be utilized to study the thermodynamic evolution of the black hole. In this work, a positive heat capacity indicates that when the temperature decreases, the black hole emits thermal heat, thus $dQ<0$ in the black hole, in order to reach thermodynamic equilibrium with the external environment, {\it i.e.}, the black hole is stable. Otherwise, a negative heat capacity represents that, if the temperature increases, the black hole also emits thermal energy toward the external environment.

A second-order phase transition is characterized by a change in the sign of the heat capacity. Due to equation \eqref{EntropiaMaster}, the sign of the derivative $\frac{\partial S}{\partial r_h}$ is positive. Furthermore, as $T>0$, the sign of the heat capacity depends solely on the sign of the derivative $\frac{\partial T}{\partial r_h}$. Consequently, for $d>4$, a phase transition occurs at $r_h=r_h^{cri}$, equation \eqref{valorCritico}, where the temperature exhibits a local maximum and the heat capacity diverges.

Below, we propose the following interpretation of the radial evolution (and consequently, the evaporation process) for both the $4D$ and extra-dimensional cases. It's worth noting that we are only providing a glimpse of the thermodynamic evolution. For a comprehensive analysis, an in-depth study of the evaporation process should be conducted in future research.

\paragraph{\bf Case $d=4$:}

As mentioned earlier, in this case, the derivative $\frac{\partial T}{\partial r_h}$ is always negative, resulting in a negative heat capacity. Consequently, when we move from right to left in Figure \ref{PlotTemperaturad4}, we can observe a simultaneous decrease in the event horizon value and an increase in temperature.

As a result, by using the relationship $C=dQ/dT$, we find that as temperature increases, the black hole emits thermal heat ($dQ<0$), while the event horizon contracts. As the event horizon contracts, it is possible to reach the value $r_h=0$, leading us to assume that the evaporation process concludes at this point. Therefore, it is conceivable that this black hole evaporates completely at a finite temperature value given by $T= \frac{1}{4\pi A}$.

This outcome differs from the Schwarzschild vacuum solution, where complete evaporation is only feasible as the temperature approaches infinity, despite the heat capacity also remaining negative. This result also differs from the evaporation process of RBHs (as seen, for example, in \cite{Estrada:2023cyx}), where it is not possible to achieve complete evaporation at $r_h=0$. 

\paragraph{\bf Extra--dimensional case :} In this case, the derivative $\frac{\partial T}{\partial r_h}$ has a negative (positive) sign at the right (left) side of $r_h=r_h^{cri}$. See figure \ref{PlotTemperaturaD}. Consequently, the heat capacity has a negative(positive) sign at the right (left) side of $r_h=r_h^{cri}$. 

Therefore, we propose the following interpretation: Starting on the right side of $r_h=r_h^{cri}$ and progressing from right to left in Figure \ref{PlotTemperaturaD}, the heat capacity is negative. Consequently, as the temperature increases and the event horizon decreases, the black hole releases thermal heat into the surrounding environment, resulting in a decrease in its thermal heat content.

Once it reaches the point where $r_h=r_h^{cri}$, there is a phase transition where the sign of the heat capacity changes from negative to positive. After this, on the left side of $r_h=r_h^{cri}$, as the temperature decreases, the black hole continues to emit thermal heat into the surrounding environment, while the value of the event horizon decreases. 

After, when the value $T=0$ is reached, the event horizon's value is zero, as shown in equation \eqref{Temperatura}. At this point, the value $C=0$ is also reached, equation \eqref{CalorEspecifico} and consequently, the emission of thermal heat into the exterior stops. Thus, the evaporation process also comes to an end. Therefore, since the process concludes at $r_h=0$, the black hole evaporates completely.

It is worth noting that to the right of $r_h=r_h^{cri}$, the specific heat is negative, just as in the vacuum Schwarzschild solution. However, the introduction of a phase change at this point and the absence of an inner horizon allows the black hole to evaporate completely down to $r_h=0$. This result also differs from regular black holes (RBHs), where complete evaporation down to $r_h=0$ is not possible because it halts at the point where the inner and event horizons coincide. So, the inclusion of our type of matter makes the evaporation process different from what occurs with the type of matter in regular black holes.

In reference \cite{Estrada:2023cyx}, it was conjectured that for higher dimensional RBHs, the evaporation would cease once the horizon radius contracts to a value close to the Planck length. This phenomenon could be linked to the emergence of quantum effects at such scales. Therefore, the Generalized Uncertainty Principle (GUP) at Planck scales should inhibit total evaporation, analogous to how the uncertainty principle prevents the collapse of the Hydrogen atom at atomic scales \cite{Adler:2001vs}. However, in Equation \eqref{valorCritico}, we can observe that for an appropriate value of the parameter $A$, the value of the critical horizon radius could be on the order of the Planck length, i.e., $r_h^{cri} \approx \ell_p$. This leads us to speculate that for our type of matter for higher dimensions, quantum effects at these scales lead to a different fate than that of an RBH. Specifically, in a higher dimensional scenario at Planck scales, a phase transition would occur that would allow for complete evaporation, unlike the aforementioned phenomena. This phase transition at Planck scales could also be speculatively associated with the effects of the GUP \cite{Li:2019pit}.

\paragraph{\bf Note:} As mentioned, the introduction of matter sources into the energy-momentum tensor in RBHs leads to the appearance of an inner horizon. The presence of this inner horizon causes the formation of a black remnant, which is understood as what remains of the black hole once the evaporation process halts at the point where the inner and event horizons coincide. In this work, we have shown that by introducing matter that leads to an integrable singularity instead of a de Sitter core located at the edge or inside the inner horizon, the evaporation process does not end in a remnant, but the black hole could evaporate completely. It is worth mentioning that, although the evaporation processes differ for the 4D and extra-dimensional cases, in both cases, complete evaporation is possible as discussed above.

\section{Energy conditions and stability}
\subsection{Energy conditions}
\begin{figure}[!h]
    \centering
    \begin{minipage}{0.5\linewidth}
        \centering
        \includegraphics[width=1\textwidth]{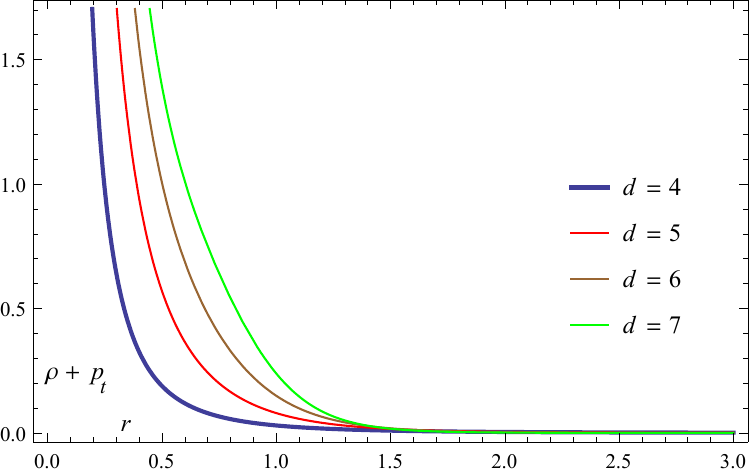}
        \label{fig:figura1minipg}
    \end{minipage}\hfill
    \begin{minipage}{0.5\linewidth}
        \centering
        \includegraphics[width=1\textwidth]{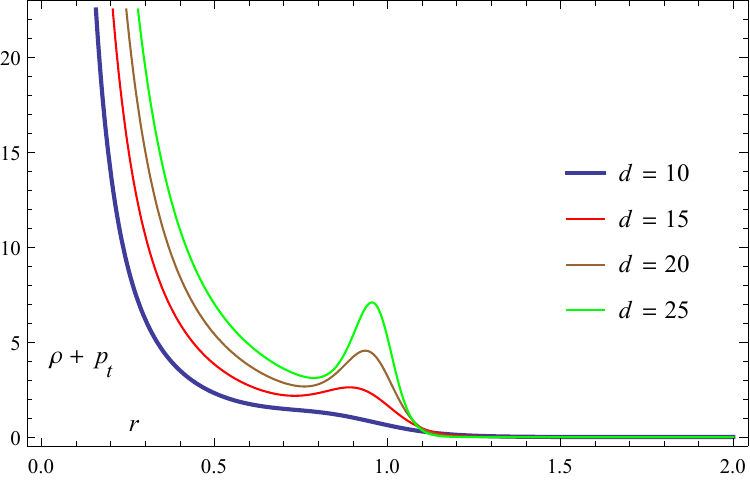}
         \label{fig:figura2minipg}
    \end{minipage}
     \caption{
Left panel: Sum of density and transversal pressure as a function of the radial coordinate, $r$, for some spacetime dimensions. Right panel: The same quantity considering higher dimensions, emphasizing the maxima near the horizon. Parameter settings: $A=M=1$.}
    \label{condenergy1}
\end{figure}

We will now examine the energy conditions associated with the source of the black hole under consideration, specifically focusing on the Weak Energy Conditions (WEC), characterized by $\rho \geq 0$ and $\rho+p_i\geq 0$, where $i$ denotes the radial ($r$) and transversal ($t$) directions. The equality trivially holds for the radial pressure, since $\rho=-p_r$ ($\rho\geq 0$, by Eq. (\ref{densidadEnergia})). Concerning the lateral pressure, we have that
\begin{equation}
   \rho+p_t= \frac{A (d-3) M \left[A+(d-2) r^{d-3}\right]}{4 \pi  r^2 \left(A+r^{d-3}\right)^3},
\end{equation}
which is positive for all $r$. In Fig. \ref{condenergy1}, we present plots of $\rho+p_t$ across various spacetime dimensions, observing thus that WEC is globally satisfied. Consequently, the Strong Energy Conditions (SEC), defined as WEC along with $\rho+\sum_i p_i\geq 0$, are also met in the entire domain.

Interestingly, a closer examination reveals that compliance with  WEC is particularly pronounced near the horizon, where a local maximum for $\rho + p_t$ is observed, especially in higher-dimensional spacetimes as depicted in the right panel. Notably, these maxima exclusively manifest in dimensions $d\geq12$. In this dimension, $\rho+p_t$ peaks at $r_{max}=\frac{A^{1/9}}{2^{1/3}}$. It is possible to demonstrate that Dominant Energy Conditions (DEC) are also always satisfied, since $\rho-|p_i|\geq 0$ for all $r$ and any values of the parameters, with this quantity presenting the same behavior as the one that governs WEC. 

Therefore, in this context and unlike the RBHs, even though the singularities are mathematically manageable (integrable), they do not violate the energy conditions established by the theory.

\subsection{Stability of the solution}

\begin{figure}[!h]
    \centering
    \begin{minipage}{0.5\linewidth}
        \centering
        \includegraphics[width=1\textwidth]{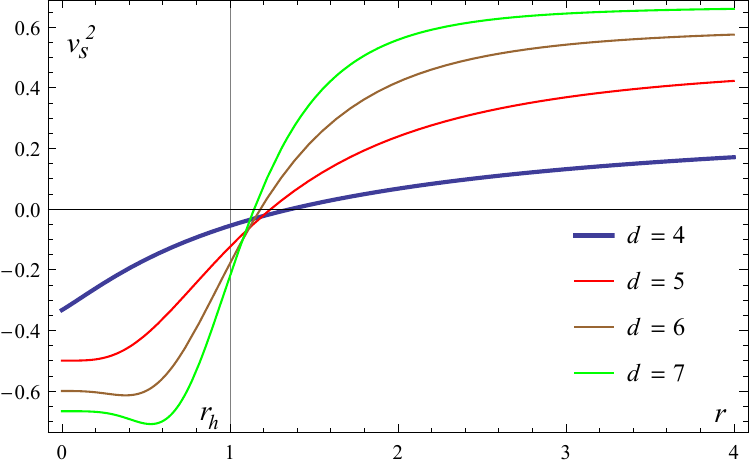}
        \label{fig:figura1minipg}
    \end{minipage}\hfill
    \begin{minipage}{0.5\linewidth}
        \centering
        \includegraphics[width=1\textwidth]{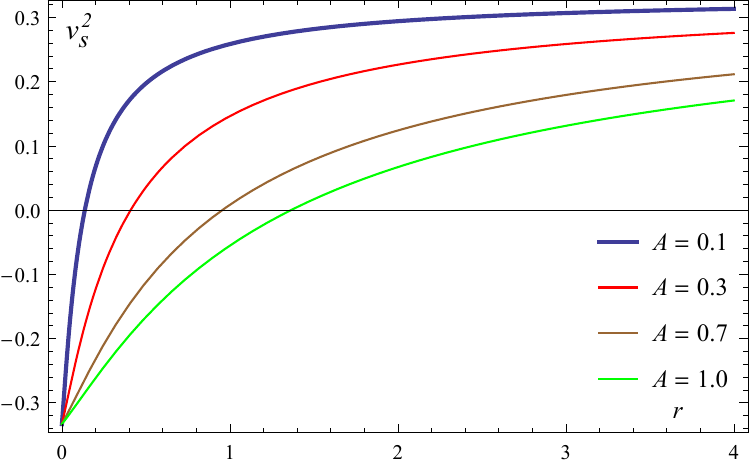}
         \label{fig:figura2minipg}
    \end{minipage}
     \caption{
Left panel: Squared sound velocity as a function of the radial coordinate, for some values of the spacetime dimension, considering $A=1$. Right panel: The same quantity for different values of the parameter $A$, taking $d=4$.}
    \label{velsound}
\end{figure}
We will now investigate the stability of the black hole solution by analyzing the fluid  sound velocity, defined by 
\begin{equation}\label{vsound}
    v_s^2=\frac{d <p>}{d\rho}=\frac{<p>'}{\rho'}=\frac{(d-3) \left[-A^2 r^6-2 A (d-3) r^{d+3}+(d-2) r^{2 d}\right]}{(d-1) \left(A r^3+r^d\right) \left[A r^3+(d-2) r^d\right]},
\end{equation}
where the average pressure $<p>$ is given by
\begin{equation}\label{averpres}
<p>=\frac{1}{d-1}[p_r+(d-2)p_t],
\end{equation}
since all lateral pressures are equal. Stability in the solution necessitates the speed of sound to adhere to $v_s \geq 0$. Additionally, it must satisfy $v_s < 1$. In the left panel of Fig. \ref{velsound}, we depict $v_s^2$ as a function of $r$ across different spacetime dimensions. Notably, stability near the horizon at $r_h = 1$ is not observed for any $d$. Achieving stability in this region (and in the remaining outer regions) requires constraints on the black hole mass, i.e.,

\begin{equation}\label{massconstraint}
    M\geq \frac{1}{2} A \left[1+\frac{1}{3-d+\sqrt{(d-5) d+7}}\right].
\end{equation}

Thus, considering $A=1$ and $d=4$, $M\gtrsim 1.18$ to one has solution stability in the vicinity of the horizon. Conversely, as it becomes represented in the right panel of Fig. \ref{velsound}, for $M=1$ and $d=4$, that stability can be observed when $A\leq 0.7$. 

\section{Radial timelike geodesics}

 With the metric in hand, it is possible to study the behavior of such space-time globally. In this context, we can study radial time-like geodesics in this geometry. By the equation \eqref{geodesica2Radial}

\begin{equation}
    V_{eff} = \frac{1}{2}\left(1 - \frac{2M}{A + r^{d-3
    }}\right)
\end{equation}

Clearly, this potential is finite for any value of $r$ and $d$, since

\begin{equation}
    \lim_{r \to 0} V_{eff} = \frac{1}{2}\left(1 - \frac{2M}{A}\right) = \mbox{finite}
\end{equation}

\begin{figure}[!h]
    \centering
    \begin{minipage}{0.5\linewidth}
        \centering
        \includegraphics[width=1.0\textwidth]{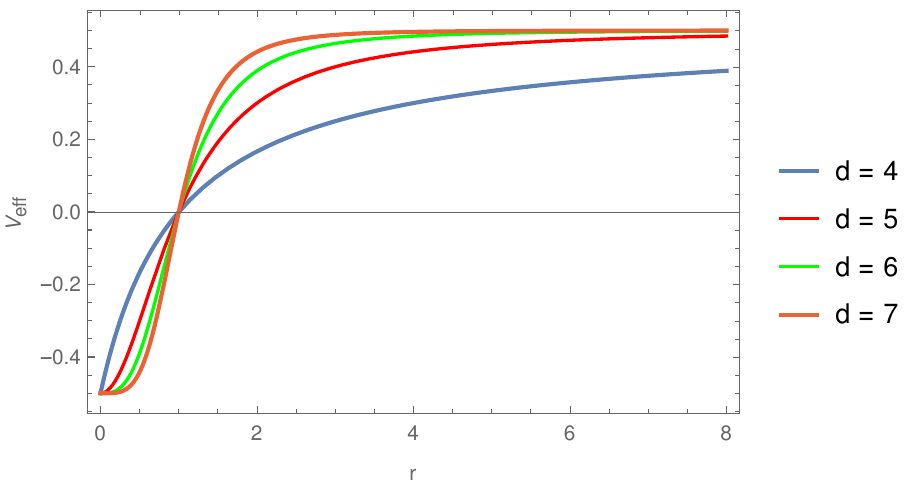}
        \label{Mplot}
    \end{minipage}\hfill
     \caption{ Effective radial potential for  $d=4,5,6,7$ with $A= M = 1$ }
    \label{V_eff_radial}
\end{figure}

In Fig. \ref{V_eff_radial} we have the plot of the effective radial potential for different values of $d$. In the figure, we can see that the potential is finite for any value of $r$ and $d$, then we can conclude that in our model the radial geodesic motion be extended up to $r = 0$.

\section{Discussion and Conclusion}

It is well known that incorporating LSM into the energy-momentum tensor leads to the emergence of an inner horizon in RBH solutions, which is linked to the physics issues outlined in the introduction. To address this problem, we have provided a $d$-dimensional black hole model with LSM in the energy-momentum tensor that lacks an inner horizon. To do this, we have established the necessary constraints to prevent the presence of such a horizon. In our model, instead of forming an unstable de Sitter core near the inner horizon, an integrable central singularity arises. Consequently, we have also determined the constraints for LSM to ensure that the volume integrals of the Ricci tensor and energy-momentum possess finite values.

Close to the origin, we have determined that the energy density must follow the form $\rho \sim r^{-2}$. This behavior in $4D$ aligns with findings near the origin documented in references \cite{Casadio:2023iqt}, which address the absence of an internal horizon, and in reference \cite{Lukash:2013ts}, which discusses finite tidal forces. Remarkably, our model guarantees finite tidal forces in the radial vicinity of such a singularity and the extendability of radial geodesics to it. Furthermore, we have shown that our type of central integrable singularity is weak in nature. Therefore, any object following a radial and timelike world-line would not be crushed by this gravitationally weak singularity

Additionally, our proposed $4D$ and higher-dimensional energy density is consistent with the speculative quantum analogy between energy density and the quantum wavefunction established for $4D$ scenarios in reference \cite{Casadio:2023iqt}. However, it is crucial to note that this aspect is only a speculative insight, and establishing a fully coherent relationship between gravitational and quantum theories necessitates further comprehensive investigation.

It is worth mentioning that the constraints for LSM enunciated in this work could serve as a recipe for constructing several $4D$ and extra-dimensional new solutions. We have constructed a full-density model that satisfies all of these constraints.

Furthermore, we have provided a glimpse into the thermodynamics evolution of our model. As mentioned, introducing matter sources into the energy-momentum tensor leads to the final stage of the evaporation process, which corresponds to a black remnant. This remnant is understood as what remains of the black hole once the evaporation process halts, at the point where the inner and event horizons coincide. It becomes impossible for a RBH to completely evaporate down to $r_h=0$. In this work, we have demonstrated that by introducing our type of matter, the evaporation process does not culminate in a remnant. Instead, the black hole could evaporate completely down to $r_h=0$, both in the $4D$ and extra-dimensional cases.

In reference \cite{Estrada:2023cyx}, it was conjectured that for higher dimensional RBHs, the evaporation would cease once the horizon radius contracts to a value close to the Planck length. This phenomenon could be linked to the emergence of quantum effects at such scales. Therefore, the Generalized Uncertainty Principle (GUP) at Planck scales should inhibit total evaporation, analogous to how the uncertainty principle prevents the collapse of the Hydrogen atom at atomic scales \cite{Adler:2001vs}. However, in Equation \eqref{valorCritico}, we can observe that for an appropriate value of the parameter $A$, the value of the critical horizon radius could be on the order of the Planck length, i.e., $r_h^{cri} \approx \ell_p$. This leads us to speculate that for our type of matter for higher dimensions, quantum effects at these scales lead to a different fate than that of an RBH. Specifically, in higher dimensional scenarios at Planck scales, a phase transition would occur that would allow for complete evaporation, unlike the aforementioned phenomena. This phase transition at Planck scales could also be associated with the effects of the GUP \cite{Li:2019pit}.

This outcome also differs from the Schwarzschild vacuum solution, where complete evaporation is only feasible as the temperature approaches infinity. In the $4D$ case, despite the heat capacity also remaining negative, it is possible for complete evaporation to occur at a finite value of temperature. In the extra-dimensional case, an additional phase transition to the Schwarzschild vacuum solution occurs, which, combined with the absence of an inner horizon, makes complete evaporation possible at zero temperature.

Our type of LSM also modifies the structure of the first law of thermodynamics, inducing a linear correction to the area law of entropy, which differs from those previously studied for RBHs \cite{Morales-Duran:2016jqt,Singh:2022xgi}. In a four-dimensional scenario, the mentioned correction is also proportional to the square of the area of a 2-sphere, $\sim \sqrt{area}$. In a speculatively way, our linear correction could be associated with quantum effects through GUP corrections in four-dimensional scenarios \cite{Majumder:2011xg,Amelino-Camelia:2005zpp}. However, this latter one requires further in-depth future work.

The fact that energy conditions can be violated for matter associated with RBH is well-known \cite{Zaslavskii:2010qz}. However, our type of matter with LSM leads to the satisfaction of energy conditions.
Additionally, our analysis of the solution stability, based on fluid sound velocity calculations, revealed that the stability near the horizon and outer regions hinge on the black hole mass satisfying the constraint outlined in Eq. (\ref{massconstraint}), varying according to the spacetime dimension.

\section*{Acknowledgements}
 Milko Estrada is funded by ANID , FONDECYT de Iniciaci\'on en Investigación 2023, Folio 11230247. Tiago M. Crispim, Geová Alencar, and Celio R. Muniz would like to thank Conselho Nacional de Desenvolvimento Científico e Tecnológico (CNPq) and Fundação Cearense de Apoio ao Desenvolvimento Científico e Tecnológico (FUNCAP) for the financial support.

\bibliography{mybib} 

\end{document}